\documentstyle[11pt,aaspp4]{article}
\tighten

\newcommand{\etal}{{\it et al.\ }}

\newcommand{\avg}[1]{\left\langle{#1}\right\rangle}

\begin{document}

\Large 

\title{Cumulant Correlators\\
       from the APM}
\normalsize 
\vskip 0.5cm 
\author{Istv\'an Szapudi$^1$ and Alexander S. Szalay$^2$} 
\vskip 1cm 
\affil{$^1$NASA/Fermilab Astrophysics Center, Fermi National Accelerator Laboratory, Batavia, IL 60510-0500}
\affil{$^2$ Johns Hopkins University, Baltimore, MD 21218 }
\vskip 0.3cm

\vskip 3cm 
\centerline{\bf Abstract} 

\small This work presents a set of new statistics, the cumulant
correlators (CC), aimed at high precision analysis of the galaxy
distribution. They form a symmetric matrix, $Q_{NM}$, related to
moment correlators the same way as cumulants are related to the
moments of the distribution.  They encode more information
than the usual cumulants, $S_N$'s, and their extraction from data
is similar to the calculation of the
two-point correlation function. Perturbation theory (PT), its
generalization, the extended perturbation theory (EPT), and the
hierarchical assumption (HA) have simple predictions for these
statistics. As an example, the factorial moment correlators measured
by Szapudi, Dalton, Efstathiou \& Szalay (1996, hereafter
\cite{sdes95}) in the APM catalog are reanalyzed using this technique.
While the previous analysis assumed hierarchical structure
constants, this method can directly investigate the validity
of HA, along with PT, and EPT.  The results in agreement with 
previous findings indicate that, at the small scales used for this
analysis, the APM data supports HA. When all non-linear corrections
are taken into account it is a good approximation at the 20 percent
level. It appears that PT, and a natural generalization of EPT for
CC does not provide such a good fit for the APM at small scales. 
Once the validity the HA is approximately established,
CCs can separate the amplitudes of different tree-types 
in the hierarchy up to fifth order.  As an example, the weights for the
fourth order tree topologies are calculated including all non-linear
corrections.

\normalsize
\vskip 0.5cm
{\bf \noindent keywords} large scale structure of the universe -- 
galaxies: statistics
-- methods: data analysis -- methods: statistical

\hfill\newpage
\section{Introduction}
Direct determination of higher order correlation functions
(\cite{fp78,peebles80}, and references therein) is burdened
with the combinatorial explosion of terms, which severely complicates
their measurement and interpretation.
Thus in the recent years indirect methods became increasingly popular for
high precision measurements of higher order correlations.  The
simplest of these methods consists of calculating the (factorial)
moments of the distribution of counts in cells, and from that, the
cumulants, $S_N$'s, of the underlying distribution
(see e.g. \cite{peebles80,gaz92,bouchet93,gaz94,cbh95,smn96}).  For a point
process, these quantities measure the amplitude of the $N$-point
correlation function averaged in a particular window.  The advantages
of this technique lie in its simplicity,
and its direct relation to the predictions of PT
(\cite{peebles80,jbc93,bern92,bern94}, EPT (\cite{cbbh96}) and
the HA (\cite{peebles80}). 
Since the averaging causes a significant loss of information,
alternative methods based on moment correlators use a pair of cells
(\cite{ssb92,mss92,sdes95}).
In the past such methods were used mainly to estimate the average
amplitude of the different $N$-point correlation
functions in the HA, the $Q_N$'s, motivated
by the theory of the BBKGY equations in the strong clustering regime.
This work presents an alternative analysis of the factorial moment
correlators which is free of assumptions,
except for the widely accepted infinitesimal Poisson model to relate
the continuum limit quantities to the measured discrete process.
Instead of fitting for the $Q_N$, a matrix $Q_{NM}$ is defined: the
CCs. Both HA and PT have specific predictions for
these possibly scale dependent quantities. After elaborating these
predictions, the method is illustrated by reanalyzing the factorial
moment correlators obtained from the APM catalog by \cite{sdes95}.  Once
the HA is established, CCs contain enough information
to separate the weights of different tree topologies up to fifth
order.  The next section outlines the basic theory, section $\S 3$
presents the predictions of PT, EPT, and HA. The measurements of the
4th order coefficients of the hierarchy from the APM catalog are
described in section $\S 4$.
\section{Theory}
Following ~\cite{sdes95} we define the factorial moment correlators
for a pair of  cells separated by a distance $r_{12}$ as
\begin{equation}
       w_{kl}(r_{12}) = { \avg{(N_1)_k (N_2)_l}-
          \avg{(N)_k}
          \avg{(N)_l}\over\avg{N}^{k+l} }, \hbox{\ \ } k \neq 0, l\neq 0,
\end{equation} 
and the normalized factorial moments for a single cell
\begin{equation}
  w_{k0} = {\avg{(N)_k}\over \avg{N}^k}.
\end{equation}
The notation $(N)_k = N(N-1)..(N-1+k)$ is introduced for the factorial
moments of the counts in cells, $\avg{}$ denotes averaging over all
cell positions in the survey.  The connection with the fluctuations
of the underlying field, $\delta$, can be obtained by formally
substituting $(N)_k/\avg{N}^k \rightarrow (1+\delta)^k$.
The generating function for the
factorial moments in terms of the cumulants $Q_N$ is
\begin{equation}
  W(x) = \exp \sum_{N=1}^\infty \Gamma_N x^N Q_N,  
\end{equation}
with
\begin{equation}
    \Gamma_N = {N^{N-2}\xi_s^{N-1}\over N!},
\end{equation}
where $\xi_s = \sigma$, the variance in a cell.
The generating function can be written in the above
form for any distribution that has cumulants. Generally, the
$Q_N$'s can have a scale dependence, while for the HA $Q_N =$ const is
expected. Note the connection with the popular alternative
notation,  $S_N = Q_N N^{N-2}$ {\it exactly}.
Similarly, the generating function of the factorial moment
correlators can be written as
\begin{equation}
     W(x,y) = W(x)W(y)\left(\exp Q(x,y)-1\right),
     \label{eq:qxy}
\end{equation}
with 
\begin{equation}
      Q(x,y)  =  \xi_l\sum_{M=1,N=1}^\infty x^My^N 
              Q_{NM}\Gamma_M\Gamma_N NM.
      \label{eq:qnm}
\end{equation}
This latter equation defines the CCs, $Q_{NM}$,
with $\xi_l = w_{11}$, the two-point correlation function
between the cells. Typically in the APM survey, $\xi_l \ll \xi_s (=\sigma) < 1$
Note that the linear dependence is factored out, however, $Q_{NM}$ is
not necessarily a constant.

Cumulants and CCs are related to the continuum limit
connected moments because of the continuum properties of the
factorial moments
\begin{eqnarray}
\frac{\avg{\delta_1^N}_c}{N!} &= Q_N \Gamma_N \\
\frac{\avg{\delta_1^N \delta_2^M}_c }{N!M!} &=
              Q_{NM}\Gamma_M\Gamma_N NM \xi_l.
\end{eqnarray}
Although the above equations are formally
identical to ~\cite{sdes95}, there are two subtle differences: there
is no reference to the hierarchical assumption, therefore $Q_{NM}$
becomes a matrix, and it is understood as an {\it exact} equation,
i.e. the non-linearities are included. It is convenient to define
CCs linear in $\xi_l$,  denoted by $\tilde Q_{NM}$,
which are obtained from the generating function with the approximation
of $\exp Q(x,y)-1 \simeq Q(x,y) + {\cal O}(\xi_l^2)$.
The $\tilde Q_{NM}$'s coincide up to 
normalization with the $C_{NM}$'s calculated from PT by \cite{bern95}
(see next section). Note that in the following linear and non-linear
always refers to powers of $\xi_l$. 

The CCs can be calculated
for any well behaved point process by
expanding $W(x,y)/[W(x)W(y)]$ according to equation~\ref{eq:qxy}.
For instance the third and fourth order moments are
\begin{eqnarray}
   Q_{12}\Gamma_1\Gamma_2 2 \xi_l =&  w_{12}/2-\xi_l \\
   Q_{13}\Gamma_1\Gamma_3 3 \xi_l =&  w_{13}/6- 
                             w_{12}/2- w_{20}/2+\xi_l \\
   Q_{22}\Gamma_2^2 4 \xi_l =&  w_{22}/4- w_{12}+\xi_l-\xi_l^2/2,
\end{eqnarray}
and follows that $Q_{22} = \tilde Q_{22} - \xi_l/2\xi_s^2$.
\section{Predictions}
In the highly nonlinear regime, the HA
(e.g.,~\cite{peebles80}; BS) states that the $N$-point correlation
functions can be written as a sum of products of $N-1$ two-point
correlation functions. Each product corresponds to a tree spanning the
$N$-points, and there is a summation over all possible trees. The
different tree topologies, labeled with $k$, are weighted with a
constant $Q_{Nk}$.  Our notation in detail can be found in 
\cite{bss94,sc96}.  
One of the goals of this paper is validate the HA to
an unprecedented accuracy.

Comparing Equation~\ref{eq:qnm} with 
~\cite{sdes95}, and~\cite{ss93a},  yields a linear order prediction 
for  the HA
\begin{equation}
   Q_{N+M} \simeq \tilde Q_{NM} \simeq const. \label{eq:qnmhier}
\end{equation}
For instance the 4th order cumulant $Q_4$ is approximately equal to
the linear CCs $\tilde Q_{13}$, $\tilde Q_{22}$, 
and constant, etc.  While form factors from the smoothing were 
shown to be negligible by
\cite{bss94}, different tree topologies and non-linear corrections
will be taken into account next for a more accurate prediction.

The only 3rd order CC is $Q_{12}$.  Tree graphs
spanning three points have only one possible topology (its weight
denoted by $Q_3$ with form factors neglected), giving altogether three
possible graphs.
\begin{equation}
  \avg{\delta_1^2\delta_2} = 2 Q_{12}\xi_s \xi_l = 
   Q_3(2 \xi_l \xi_s + \xi_l^2), \label{eq:q3nl}
\end{equation}
reproducing $\tilde Q_{12} = Q_3$ at linear order.

At fourth order there are two CCs $Q_{13}$, and
$Q_{22}$. The sixteen possible trees spanning four points come in two
distinct topologies: four ``snake'' graphs and twelve ``star'' graphs.
Their respective amplitudes are denoted with $R_a$ and $R_b$ in the HA.
Summing all possible graphs with the appropriate statistical weights
gives
\begin{equation}
  \avg{\delta_1^3 \delta_2}_c = 9 Q_{13} \xi_l\xi_s^2 =
  6 \xi_l \xi_s^2 R_a+ 
  3 \xi_l \xi_s^2 R_b + 
  6\xi_l^2 \xi_s R_a +
  \xi_l^3 R_b, \label{eq:rarbnl1}
\end{equation}
and \begin{equation}
  \avg{\delta_1^2 \delta_2^2}_c = 4 Q_{22} \xi_l\xi_s^2 =
  4 \xi_l \xi_s^2 R_a+ 
  4 \xi_l^2 \xi_s R_a + 
  4\xi_l^2 \xi_s R_b +
  4 \xi_l^3 R_a. \label{eq:rarbnl2}
\end{equation}
These two equations are linear in $R_a$ and $R_b$, therefore they can
be solved yielding equations (with non-linear coefficients in terms of
$\xi$) in terms of $Q_{13}$ and $Q_{22}$.  The linear
solution is $R_a = \tilde Q_{22}$ and $R_b = 3\tilde Q_{13}-2\tilde
Q_{22}$.


Direct comparison of Equation ~\ref{eq:qnm} with  the coefficients $C_{NM}$
in ~\cite{bern95} reveals that they are identical to the 
linear order CCs up to normalization
\begin{equation}
   C_{NM}  = \tilde Q_{NM} N^{N-1} M^{M-1} + {\cal O}(\xi_l^2).
\end{equation}
Perturbation theory predicts that the coefficients
factorize such that
\begin{equation}
  C_{NM} = C_{N1} C_{M1} 
  \label{eq:cnmfact},
\end{equation}
and the series $C_{N1}$ was calculated up to first non-trivial
order. 
The interested reader is referred to 
\cite{bern95} for detailed predictions in the weakly
non-linear regime, for the present work only Equation~\ref{eq:cnmfact} 
is needed.

Although biasing is not investigated in this paper,
it is worth to note that it can significantly change the higher
order correlations. In the weakly non-linear regime the results of
\cite{fg93} should be generalized for CCs. 
Such a  calculation, which is left for subsequent research,
will resolve the remaining ambiguities in the interpretation
of CCs.
\section{Measurements from the APM Catalog}
For an initial assessment, the linear CCs were first
calculated from the factorial moment correlators measured 
in the APM survey
(\cite{mad90a,mad90b,mad90c}) by~\cite{sdes95}.  
In what follows, a density map of cell
size $0.23^\circ$ and magnitude cut of $b_J = 17-20$ was used
(see \cite{sdes95} for the detailed properties of the density maps).
The bottom panel of the Figure shows 
the measured $\tilde q_{NM}$'s (the linear projected
CCs; lower case symbols refer to projected
quantities) up to fifth order. To interpret the figures note
that the CCs are characterized by
two relevant scales: the angular
separation, and the smoothing scale, or cell size.
On the figures, only the separation is
shown in degrees, ($1^\circ\simeq 7 h^{-1}Mpc$ for this
magnitude cut), while the smoothing length 
(always $0.23^\circ$) remains implicit.
The degeneracy and the approximate
parallel nature of the curves immediately suggest that the HA is a
reasonable approximation. 
At larger scales the CCs appear to roll off,
while the prediction stays flat, and the degeneracy of the curves is
slightly broken. This is mainly due to fact that linear CCs were used, and
cumulants are not exactly constant at all scales as shown by
\cite{gaz94,smn96} (i.e. HA is slightly broken).  

The middle panel of the Figure. illustrates equation~\ref{eq:cnmfact}
predicted by leading order PT.  The solid lines are the CCs
$\tilde q_{NM}, N,M > 1$, while the dotted lines show the
corresponding $\tilde q_{1N}\tilde q_{1M}$.  Only the fourth and fifth
order are shown. The degree of validity of PT 
can be judged from how well the dotted and solid lines match.  Since the
dotted lines appear to be consistently smaller than the solid ones 
this model provides a less
accurate description of the data than HA. Possibly, higher 
than leading order PT
could improve the representation of the data; it is left for future
work.  

It can be argued, that PT for the CCs is valid when both
relevant scales are in the weakly non-linear regime.
While PT matches the higher order correlations in the APM
for larger scales (\cite{gf94}),
for the small cell size used in this work
non-linearities can be important for the present measurement
(\cite{bg96}). However,
it was found in $N$-body simulations
(\cite{cbbh96}), and galaxy data (\cite{smn96}),
that the higher order correlation
amplitudes, $Q_N$, measured from counts in cells are similar to
the one prescribed by PT, but with a steeper power spectrum.
This phenomenological extension of PT is the essence of EPT.
The previous exercise taken at face value would suggest
that EPT cannot be generalized for moment correlators. 
A rough estimate of the errors based on Equation
\ref{eq:qnmhier} with scaling the variance from 
\cite{gaz94}, and Gazta\~naga 1996 (private 
communication)  yields 5\%, 7\%, and 7\% for the third, fourth, and
fifth order respectively. These error-bars, which are not
necessary conservative, could only marginally exclude 
the natural extension of EPT at small scales.
Further measurements in $N$-body simulations, and high quality data
are needed to show, whether the EPT paradigm can be applied to CCs.


The HA can be examined with further scrutiny by relaxing
the previous assumptions on linearity and uniform weighting
of topologies. The form factors resulting from the pair of
cells are expected to be smaller than the measurement errors and will
be still neglected. Counting the number of degrees of freedom reveals
that from the cumulants and CCs it is possible to
separate the different tree topologies up to fifth order. 
A calculation for the third and fourth order is presented here. The fifth
order calculation is analogous, although somewhat tedious. At higher
than fifth order additional information is needed to separate the
different graph types.

The long dashed line on the top panel of the Figure shows
the non-linear measurement of $q_3$ as calculated from $q_{21}$
of the APM according to Eq.~\ref{eq:q3nl}.
The dotted lines show the linear solution $r_a$, and $r_b$ as
computed from $\tilde q_{22}$, and $\tilde q_{31}$.
The hierarchy
predicts two horizontal lines, with the constraint that $ 16 q_3 = 12
r_a + 4 r_b $. The linear approximations on the other hand show a
strong scale dependence, increasing and even crossing over at
the smallest scales: a possible sign of non-linear effects.  The full
non-linear equations (\ref{eq:rarbnl1}, \ref{eq:rarbnl2}) yield the result
plotted with solid lines: the non-linear corrections remove most of
the scale dependence, as expected if HA is satisfied.
 The residuals are probably due to the neglected
form factors, measurement errors. On the left side of the panel
several amplitudes are plotted for comparison; for these points
the angular scale is irrelevant.
The three sided symbols refer to third order quantities, the 
four sided to fourth order. The filled triangles and squares
shows the value of $q_3 = 1.15$ and $q_4 = 2.2$ calculated from the averaged
value of $q_{21} = 1.15$, and $r_a = 1.15$, and $r_b = 5.3$, respectively. 
The open symbols correspond to the values
of $q_3 = 1.7$, and $q_4 = 4.17$ measured from the factorial moments
alone, $w_{k0}$, at the scale of the cells. For a comparison, the two stars
show the respective measurements of \cite{sdes95}
 $q_3 = 1.16$, and $q_4 = 1.96$. 
The reason that \cite{sdes95}
measured a somewhat lower $q_4$ is that they used 
linear approximations (dotted lines) only.
The measurements of 
$q_3 = 1.7 $ by \cite{gaz94} in the APM and
$q_3 = 1.6$ \cite{smn96} at the same cell size, are 
in excellent agreement with the results from $w_{k0}$.
The values
for the fourth order in the same sources, $q_4= 3.7$ and $q_4 = 3.2$, 
are slightly lower than above, but the agreement is still within
$20-30$ percent.

The above numbers suggest that, while 
the different measurements using the same method are consistent
with each other even in different catalogs,
there is some disagreement between the results based
on moment correlators and moments.
The error distribution studied by \cite{sc96}
provides useful clues to  resolve this apparent discrepancy. 
Since the distribution of errors is positively skewed and increasingly
so for higher order moments, an upward fluctuation is more likely
than a downward.
This effect is increasing with the order of the moments measured.
In the 
method proposed by this work $q_3$ is estimated from the value of $q_{21}$.
The behavior of the errors is similar to the multiple
of a second and first order quantity, thus the variance
is reduced. Note that this is possible, only after the
hierarchy is established, i.e. a prior information is used to
reduce the scatter from cosmic errors.
An accurate error estimation in this case would involve a tedious
calculation, a non-trivial generalization of \cite{sc96}. 

The de-projection using the coefficients in \cite{sdes95} yields
$Q_3 = 1$, $R_a = 0.8$, and $R_b$ = 3.7, giving $Q_4 = 1.5$. 
This is to be compared with
with \cite{fp78}, where the direct determination of the four-point
correlation function from the Lick catalog yielded $R_a = 2.5 \pm 0.6$
and $R_b = 4.3 \pm 1.2$. 
These results could give a clue for solving the BBKGY equations
in the highly non-linear regime.
The assumption of \cite{ham88}, that only the snake graphs
have a contribution, appears to be close to our results: although both graph
types have a contribution, the average is closer to the snake coefficient.
The ansatz of \cite{bs92}, $\sqrt{R_a} \simeq Q_3$, is
not a particularly good approximation.
In conclusion the statistics of the CCs
is in excellent agreement with HA. The method outlined here
in conjunction
with future data and $N$-body simulations will be able to pin
down the amplitudes of the higher order correlations with
unprecedented accuracy.
\acknowledgments
We would like to acknowledge discussions with F. Bernardeau,
S. Colombi, and J. Frieman, and suggested improvements by the referee, 
E. Gazta\~naga.
The original measurement of the
factorial moment correlators was carried out in collaboration
with G. Dalton, and G. Efstathiou.
I.S. was supported by DOE and NASA through grant
NAG-5-2788 at Fermilab. A.S.S. was supported by a NASA LTSA grant.

\hfill\newpage

\section{Figure Caption}

\noindent Lower Panel. The linear CCs, $\tilde q_{nm}$,
the main raw results of the paper are displayed up to fifth order
as a function of the angular separation of cells in degrees.
The parallel degenerate lines suggest the HA.

\noindent Middle Panel. 
The linear CCs are shown
on a linear scale (solid lines) together with the prediction from
PT (dotted line). The agreement is improving towards the higher
scales.

\noindent Upper Panel. 
The hierarchical amplitudes as calculated
from the fully non-linear CCs are displayed.
The long dashed line corresponds to the estimator of $q_3$, 
the solid lines to the estimator of $r_a$, and $r_b$, the amplitudes
of the fourth order snake, and star graphs, respectively. The dotted
lines show the linear approximation, which breaks down at smaller
scales at this level of precision. The filled symbols mark
$q_3$ (triangle), and $q_4$ (square) as calculated from the
moment correlators. The open symbols are the same as measured
from the moments of counts in cells only. Finally, the crosses
show the measurements of $q_3$ (triangular), and $q_4$ (square)
by \cite{sdes95} for comparison.

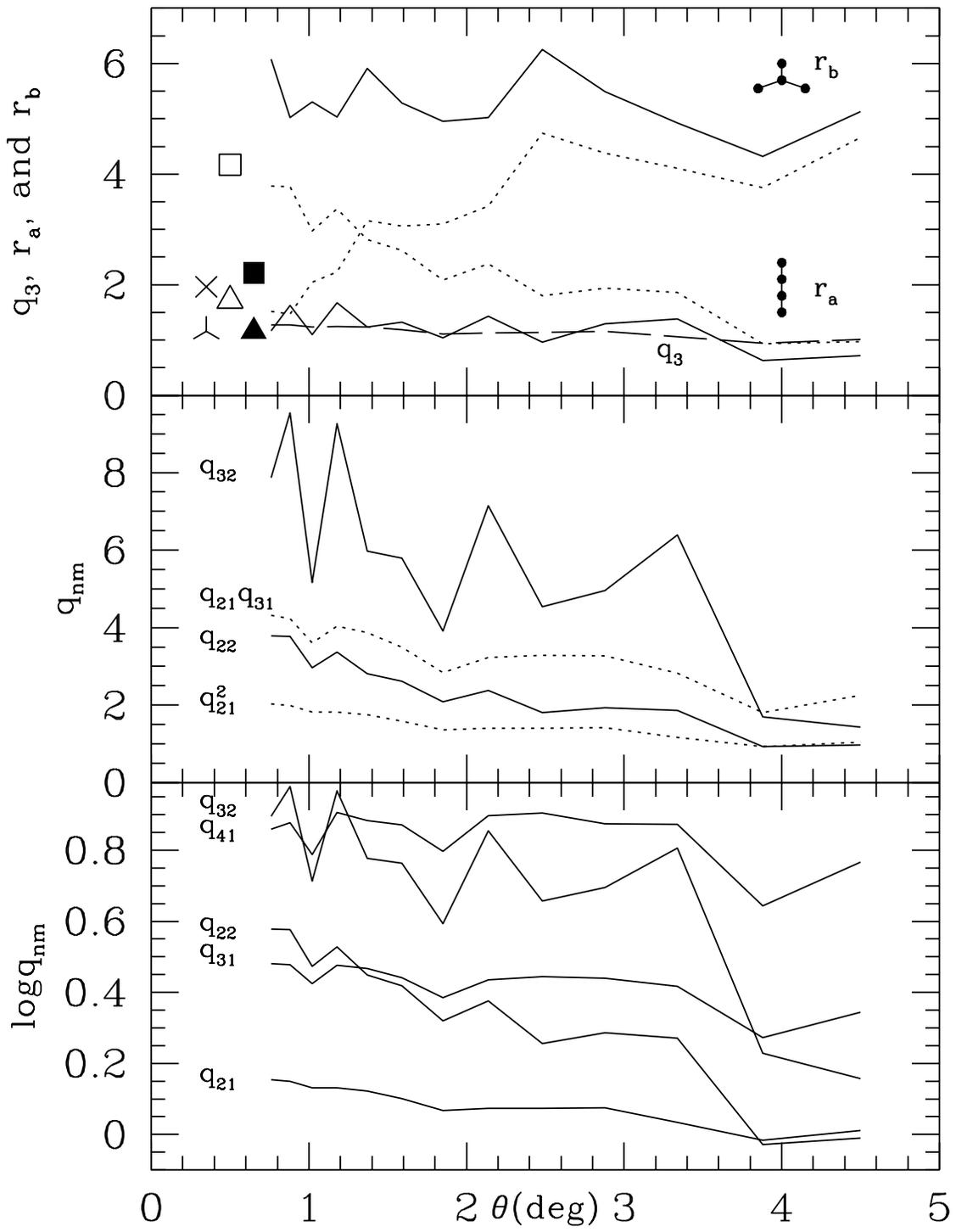
\end{document}